\documentclass[epj]{webofc}
\usepackage[varg]{txfonts}   
\usepackage{graphicx,color} 
\usepackage{bm}       
\usepackage{amsmath}  
\usepackage{amssymb}

\newcommand {\nc} {\newcommand}

\newcommand{\Ref}[1]{Ref.~\cite{#1}}
\newcommand{\Eq}[1]{Eq.~\eqref{#1}}

\newcommand{\real}{\,\mathrm{Re}\,}
\newcommand{\imag}{\,\mathrm{Im}\,}
\nc {\IR} [1]{\textcolor{red}{#1}}
\nc {\IB} [1]{\textcolor{blue}{#1}}
\nc {\IP} [1]{\textcolor{magenta}{#1}}
\woctitle{21st International Conference on Few-Body Problems in Physics}
\begin{document}
\title{
Towards a Faddeev-AGS description of $(d,p)$ reactions
with heavy nuclei: Regularizing integrals with Coulomb functions.}
\author{V.~Eremenko\inst{1,5}\fnsep\thanks{\email{to@vsl.name}} \and
        L.~Hlophe\inst{1}\fnsep \and
        Ch.~Elster\inst{1}\fnsep\thanks{\email{elster@ohio.edu}} \and
        F.~M.~Nunes\inst{2}\fnsep \and
        I.~J.~Thompson\inst{3}\fnsep \and
        G.~Arbanas\inst{4}\fnsep \and
        J.~E.~Escher\inst{3}\\[1mm]
        \textbf{TORUS} Collaboration${}^\dagger$
        (\url{http://reactiontheory.org})
}

\institute{INPP
  \textit{and}
  Department of Physics and Astronomy,
  Ohio University, Athens, Ohio 45701, USA
\and
  NSCL
  and
  Dept. of Physics and Astronomy,
  Michigan State University, East Lansing, Michigan 48824, USA
\and
  Lawrence Livermore National Laboratory,
  L-414, Livermore, California 94551, USA
\and
  Reactor and Nuclear Systems Division,
  Oak Ridge National Laboratory, Oak Ridge, TN 37831, USA
\and
  SINP
  M.V. Lomonosov Moscow State University, Moscow, 119991, Russia
}

\abstract{
The repulsive Coulomb force poses severe challenges when describing
 $(d, p)$ reactions for highly charged nuclei as a three-body
problem.
Casting Faddeev-AGS equations in a Coulomb basis avoids introducing screening of the
Coulomb force.
However, momentum space partial-wave $t$-matrix elements need to be evaluated in this
basis. When those $t$-matrices are separable, the evaluation  
requires the folding
of a form factor, depending on one momentum variable,
with a momentum space partial-wave Coulomb function,
which has a singular behavior at the
external momentum $q$.
We developed an improved regularization scheme
to calculate Coulomb distorted
form factors as the integral over the Coulomb function and complex
nuclear form factors. 
}
\maketitle
%

\section{Introduction}\label{intro}
Direct reactions involving rare isotopes offer an important tool for
understanding the structure of such nuclei.
Deuteron induced reactions are attractive from the experimental
perspective, since deuterated targets are readily available.
Theoretically they are attractive, since the scattering problem
can be reduced to an effective three-body problem,
which can be solved exactly using Faddeev techniques.
The momentum space Faddeev equations (here in the Alt-Grassberger-Sandhas
form) have successfully been solved for $(d, p)$ reactions involving
light nuclei~\cite{Deltuva2009}.
However, the screening technique employed to handle the Coulomb
force encounters technical difficulties, when applied to
$(d, p)$ reactions with heavier nuclei~\cite{Upadhyay2012}.
An alternative to the screening procedure is a solution of
the Faddeev-AGS equations in the Coulomb basis.
This was suggested and carried out in \Ref{Mukhamedzhanov2012}
using real two-body transition operators in separable form,
neglecting spin degrees of freedom. 

Casting Faddeev-AGS equations in the Coulomb basis requires the evaluation
of momentum-space partial-wave matrix elements of two-body transition
operators,
\begin{equation}
  t^{C}_{l}(q', q, E)
  =
  \int dp' \, dp \,
  \psi^{C}_{l, q', \eta'}(p')^\dagger
  \, t_{l}(p', p, E)
  \, \psi^{C}_{l, q, \eta}(p).
\end{equation}
If $t_l(E)$ is separable, one needs to consider
\begin{equation}
  t^C_l(q, q', E) =
  \sum_{zy} u^C_{l, z}(q) \, \lambda_{l,zy}(E) \, u^C_{l, y}(q')^\dagger.
\end{equation}
In this case only  the folding
of a form factor $u_l(p)$, depending on one momentum variable,
with partial-wave Coulomb function $\psi^C_{l, q}(p)$, for which we developed the
numerical procedures in~\Ref{Eremenko2015},  is required:
\begin{align}
  u^C_{l, z}(q) &=
    \int \frac{dp \, {p}^2}{2 \pi^2}
    u_{l, z}(p)
    \psi^C_{l, q, \eta}(p)^*,&
  \text{and}& &
  u^C_{l, y}(q')^\dagger &=
    \int \frac{dp \, {p}^2}{2 \pi^2}
    u_{l, y}(p)
    \psi^C_{l, q', \eta}(p).\label{Eq:CoulDistFF}
\end{align}
Both integrals have  an oscillatory singularity in the point $p = q$,
where $q$ is the external momentum. The partial-wave Coulomb functions are
given by
\begin{align}
  \psi^C_{l, q, \eta}(p) \overset{p \to q}{=}
    \psi^C_{l, q, \eta}(\chi) &\overset{\chi \to 0}{=}
    \mathcal{A}(q, l, \eta)
    \left[
      \frac{\mathcal{B}(\chi, q, l, \eta)}{(\chi + i 0)^{1 + i \eta}}
      -
      \frac{\mathcal{B}(\chi, q, l, \eta)^*}{(\chi - i 0)^{1 - i \eta}}
    \right],&
  \chi &\equiv p - q.\label{Eq:CoulombFunction}
\end{align}
To evaluate $u^C_{l, z} (q)$, one needs to regularize the integral
in \Eq{Eq:CoulDistFF}.
In \Ref{Mukhamedzhanov2012} this regularization is performed by
using a Gel'fand-Shilov (see \Ref{GelfandShilov1964}) technique
for real form factors. When working with form factors describing
complex nucleon-nucleus potentials, the
regularization scheme must be constructed for  complex
form factors.

\section{Regularization Procedure}\label{Sect:Regularization}

Here we present a regularization scheme for calculating Coulomb distorted
form factors as integrals over a Coulomb function and a complex
form factor, which differs from the scheme
presented in~\Ref{Upadhyay2014}.
The key point of the work of~\Ref{Upadhyay2014} 
is that the integral is  regularized  only in a
tiny neighborhood ($\Delta \sim 10^{-6}$~fm${}^{-1}$) around $q$.
That allows that the regularized singular part of the folding integral
can be reduced to an analytic expression.
However, the non-regularized parts of the integral
must be calculated quite close to the value of $q$,
which requires a lot of integration mesh points independent of the numerical
integration quadrature used.
For example, we used thousands of Gauss-Legendre quadrature
points to obtain results converged to 3 significant figures. 

\begin{figure}[tb]
\centering
\sidecaption
\includegraphics[height=5cm,clip]{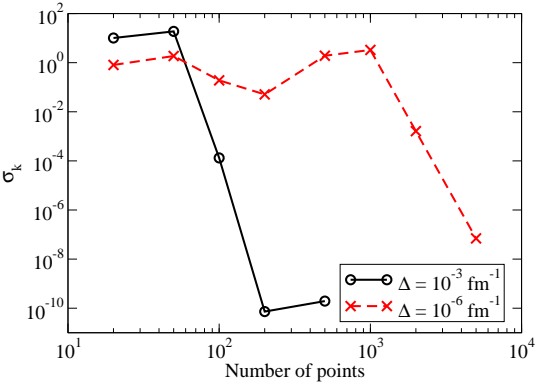}
\caption{The error plot of the non-regularized integrals $I_a^b$
with the same lower limit $a = 0.4$~fm${}^{-1}$,
but different upper limits:
[solid black]~$b = 0.499$~fm${}^{-1}$
  (i.e. $\Delta = 1 \cdot 10^{-3}$~fm${}^{-1}$),
[dashed red]~$b = 0.499999$~fm${}^{-1}$
  ($\Delta = 1 \cdot 10^{-6}$~fm${}^{-1}$).
The parameters are: $q = 0.5$~fm${}^{-1}$, $l = 0$, $\eta = 5.3$.
These values of $q$ and $\eta$ correspond to the $p + {}^{208}\text{Pb}$
system with $E_{c.m.} = 6$~MeV.
See Sects.~\ref{Sect:Regularization} and~\ref{Sect:Results}
for discussion and definition of the relative error $\sigma_k$.
}\label{fig:BlueErrorplot}
\end{figure}

To avoid using an excessive number of integration points, we developed here the new version
of the regularization scheme, where we apply the regularization
in a small, but finite neighborhood
($\Delta \sim 10^{-3}$~fm${}^{-1}$) around $q$.
Now the non-regularized integrals are converging much faster to an even better accuracy
(see Fig.~\ref{fig:BlueErrorplot}),
while the regularized part is still mostly dominated by the
analytical terms.
Despite the fact that we need to calculate the regularized integral,
we do not need to calculate it as accurately, as we had to calculate
the non-regularized parts in~\Ref{Upadhyay2014}.  
Thus, the required amount of CPU time is greatly reduced.

\noindent
Similar to~\Ref{Upadhyay2014}, we start by splitting the
folding integral into  regularized
and non-regularized parts,
\begin{align}
  u^C_{l, y}(q')^\dagger &=
    I_0^{q' - \Delta} + I_{q' - \Delta}^{q' + \Delta} + I_{q' + \Delta}^{\infty},&
  I_a^b &=
    \int_a^b \frac{dp \, {p}^2}{2 \pi^2}
    u_{l, y}(p)
    \psi^C_{l, q', \eta}(p).\label{Eq:IntegralAB}
\end{align}
We first focus on the regularized part $I_{q - \Delta}^{q + \Delta}$,
which we rewrite by substituting \Eq{Eq:CoulombFunction} into
\Eq{Eq:IntegralAB} as
\begin{align}
I_{q' - \Delta}^{q' + \Delta} &\equiv
  \mathcal{I}_{-\Delta}^{\Delta}
  = \frac{\mathcal{A}(q', l, \eta)}{2 \pi^2}
  \left(
  \mathcal{I}_+ - \mathcal{I}_-
  \right),&
\mathcal{I}_\pm &\equiv
  \int_{-\Delta}^{\Delta} d\chi
  \frac{\phi_\pm(\chi, q', l, \eta)
  }{(\chi \pm i 0)^{1 \pm i \eta}},\label{Eq:ToBeRegularized}
\end{align}
where
\begin{align}
\phi_+(\chi, q', l, \eta) &\equiv
  u_{l, y}(q' + \chi) (q' + \chi)^2 \mathcal{B}(\chi, q', l, \eta),\\
\phi_-(\chi, q', l, \eta) &\equiv
  u_{l, y}(q' + \chi) (q' + \chi)^2 \mathcal{B}(\chi, q', l, \eta)^*.
\end{align}
To apply the regularization scheme, we split the integrals
$\mathcal{I}_\pm$ and take the limit $\pm i 0$,
\begin{equation}
\mathcal{I}_\pm =
  \int_{0}^{\Delta} d\chi
  \frac{\phi_\pm(\chi, q', l, \eta)}{\chi^{1 \pm i \eta}}
  +
  e^{\pi \eta} \int_{0}^{\Delta} d\chi
  \frac{\phi_\pm(-\chi, q', l, \eta)}{\chi^{1 \pm i \eta}}.
\end{equation}
While taking the limit, we used the following property,
\begin{equation}
(\chi \pm i 0)^{-1 \mp i \eta}
= |\chi|^{-1 \mp i \eta} (e^{\pm i \pi})^{-1 \mp i \eta}
= -e^{\pi \eta} |\chi|^{-1 \mp i \eta}.
\end{equation}
By applying the Gel'fand-Shilov regularization technique
from~\Ref{GelfandShilov1964},
we obtain
\begin{multline}
\mathcal{I}_\pm
  =
  \int_{0}^{\Delta} d\chi
  \frac{\widetilde{\phi}_\pm(\chi, q', l, \eta)}{\chi^{1 \pm i \eta}}
  +
  e^{\pi \eta} \int_{0}^{\Delta} d\chi
  \frac{\widetilde{\phi}_\pm(-\chi, q', l, \eta)}{\chi^{1 \pm i \eta}}
\\
\mp
  \frac{\phi_\pm(0, q', l, \eta)}{i \eta}
  \left(
    1 - e^{\pi \eta}
  \right)
  \Delta^{\mp i \eta}
  +
  \frac{\phi'_\pm(0, q', l, \eta)}{1 \mp i \eta}
  \left(
    1 + e^{\pi \eta}
  \right)
  \Delta^{1 \mp i \eta},\label{Eq:RegularizedIntegral}
\end{multline}
where
\begin{align}
\widetilde{\phi}_\pm(\chi, q', l, \eta) &\equiv
  \phi_\pm(\chi, q', l, \eta)
  -
  \phi_\pm(0, q', l, \eta)
  -
  \phi'_\pm(0, q', l, \eta) \chi,\\
\phi'_\pm(0, q', l, \eta) &=
  \left.
  \frac{d\phi_\pm(\chi, q', l, \eta)}{d\chi}
  \right|_{\chi = 0}.
\end{align}
To calculate $\phi'_\pm(0, q', l, \eta)$,
we also need $du_{l, y}(p) / dp$.
Taking the limit $\Delta \to +0$, one arrives at the expressions
from~\Ref{Upadhyay2014}.

By using the expressions from~\Eq{Eq:RegularizedIntegral} and
\Eq{Eq:ToBeRegularized}, we can compute the regularized part
$I_{q - \Delta}^{q + \Delta}$ for any given finite $\Delta$.
Then, using~\Eq{Eq:IntegralAB} we evaluate
$u^C_{l, y}(q')^\dagger$,
while $u^C_{l, z}(q)$ is given by
\begin{equation}
u^C_{l, z}(q) =
\left[
\int \frac{dp \, {p}^2}{2 \pi^2}
u_{l, z}(p)^*
\psi^C_{l, q, \eta}(p)
\right]^*.
\end{equation}

To calculate  $\phi'_\pm(0, q', l, \eta)$,  the expression
for $d\mathcal{B} / d\chi$ is required.
By using Eq.~(B9) from~\Ref{Upadhyay2014},
since $p = q' + \chi$,
Eq.~(15.5.1) from \Ref{DLMF}, and by utilizing the property
of the hypergeometric function ${}_2F_1(\ldots; 0) = 1$,
we obtain
\begin{equation}
\left. \frac{d \mathcal{B}(\chi, q, l, \eta)}{d\chi} \right|_{\chi = 0}
=
e^{-i \sigma_l} \Gamma(1 {+} i \eta)
(-3 + i \eta) 2^{2 l - 2 + i \eta}
q^{l - 3 + i \eta}.
\end{equation}
Here we substituted $q'$ by $q$ to simplify the expression.

\section{Results, Discussions and Outlook}\label{Sect:Results}

To prevent the unreasonably large computational times for calculating 
Coulomb distorted form factors,
we improved the regularization scheme of~\Ref{Upadhyay2014}.
This new scheme converges much faster,
as the number of integration points increases.
To demonstrate it, we present the error plot on
Fig.~\ref{fig:BlueErrorplot}.
The integral $I_a^b$ (see caption for $a$ and $b$)
was calculated by using Gauss-Legendre quadrature
with 10, 20, 50, 100, 200, 500, 1~000, 2~000, and 5~000 points.
Starting from the second step ($k = 2$, i.e. 20 points),
the relative discrepancy
was calculated,
\begin{equation}
\sigma_{k} \equiv
\left. \left(
  \frac{|\real(I_a^b)_k - \real(I_a^b)_{k - 1}|}{|\real(I_a^b)_k|}
  +
  \frac{|\imag(I_a^b)_k - \imag(I_a^b)_{k - 1}|}{|\imag(I_a^b)_k|}
\right) \right/ 2,
\end{equation}
where $(I_a^b)_k$ is the value of $I_a^b$, obtained on the $k$-th step.
For the sake of illustration, here we were using simple
Yamaguchi-style real form factor with
constant 2.7~fm${}^{-1}$.
It should be noted, that for this large value of $\eta = 5.3$
and small value of
$q = 0.5$~fm${}^{-1}$
(the corresponding system is $p + {}^{208}$Pb at $E_{c.m.} = 6$~MeV)
the final result is being calculated
by subtraction of two large numbers,
and the integral $I_a^b$
must be computed with $\sigma_k \lesssim 10^{-4}$ to get even
the first digit of the final result. This new regularization scheme will be essential
for successful progress in the numerical implementation 
of the Faddeev-AGS equations
in Coulomb basis including spin degrees of freedom.

\begin{acknowledgement}
\vspace{-3mm}
This work was performed in part under the auspices of the
US Department of Energy, Office of Science of Nuclear Physics,
under the topical collaborations in nuclear theory program
No.~DE-SC0004087 (TORUS Collaboration), and under Contract DE-FG02-93ER40756 with Ohio
University.
\end{acknowledgement}

%
%
%

\end{document}